\documentclass[aps,prd,twocolumn,tightenlines,nofootinbib,superscriptaddress,floatfix]{revtex4-2}
 
\usepackage{slashed}
\usepackage{hyperref}       
\usepackage{amsmath}
\usepackage{feynmp-auto} 
\usepackage{graphicx} 
\usepackage{xcolor}
\definecolor{unicolor}{RGB}{193,11,37}

\begin{document}

\title{Sensitivity of MAGIX@MESA to BSM effects via Bethe-Heitler pair production}

\author{Aleksandr Pustyntsev}
\affiliation{Institut f\"ur Kernphysik and $\text{PRISMA}^{++}$ Cluster of Excellence, Johannes Gutenberg Universit\"at, D-55099 Mainz, Germany}
\author{Marc Vanderhaeghen}
\affiliation{Institut f\"ur Kernphysik and $\text{PRISMA}^{++}$ Cluster of Excellence, Johannes Gutenberg Universit\"at, D-55099 Mainz, Germany}

\date{\today}

\begin{abstract}
We explore the sensitivity of the upcoming MAGIX experiment at the MESA facility to light Beyond the Standard Model (BSM) mediators in the few to hundred MeV mass range. Utilizing high-intensity electron beams of 55 MeV and 105 MeV on a heavy $^{181}\text{Ta}$ target, we investigate the production of scalar, pseudoscalar, vector, and axial vector mediators via the Bethe-Heitler process. By optimizing the asymmetric kinematic acceptance of the double-spectrometer setup to enhance the signal over background ratio, we demonstrate that MAGIX can probe mediator-electron couplings down to $\mathcal{O}(10^{-4})$, offering a competitive probe of the dark sector in the sub-GeV mass range.
\end{abstract}

\maketitle

\section{Introduction}\label{sec2}

While the Standard Model (SM) provides a highly successful description of fundamental interactions, phenomena such as dark matter \cite{Bertone:2004pz,Planck:2018vyg} and persisting experimental anomalies—such as the X17 \cite{Krasznahorkay:2015iga,Krasznahorkay:2021joi,PADME:2025dla}—strongly motivate the search for BSM physics. A compelling paradigm posits a “dark sector” populated by light, weakly coupled degrees of freedom, such as dark photons or axion-like particles (ALPs) \cite{Holdom:1985ag,Pospelov:2007mp,Ringwald:2012hr,Baker:2013zta}. These mediators, typically residing in the MeV to GeV mass range, interact with the visible sector via renormalizable portal operators or as pseudo-Nambu-Goldstone bosons \cite{Bauer:2017ris,Knapen:2017xzo,Fabbrichesi:2020wbt}. 

Given the absence of New Physics signals at high-energy colliders, low-energy precision experiments are playing an increasingly important role in exploring the dark sector parameter space \cite{Jaeckel:2010ni}. Among these are the PADME experiment \cite{PADME:2025dla}, the Jefferson Lab polarized positron program \cite{Accardi:2020swt}, and the upcoming Mainz Energy-recovering Superconducting Accelerator (MESA) facility—specifically, the MAinz Gas Internal target eXperiment (MAGIX) therein \cite{Schlimme:2024eky}.

The latter is the main subject of this work. In this paper, we investigate the sensitivity of the MAGIX@MESA experiment to visibly decaying dark sector mediators during its initial extracted-beam phase. By analyzing the QED-induced Bethe-Heitler background and the BSM cross section contributions for generic scalar, pseudoscalar, vector, and axial vector mediators, decaying into an $e^- e^+$ pair, we present projected exclusion limits that demonstrate the experiment's capability to probe newly accessible regions of the BSM parameter space.

The paper is organized as follows. In Section \ref{sec2}, we outline the MAGIX@MESA experimental setup and kinematics configurations utilized in this study. Section \ref{sec3} details the Bethe-Heitler pair production mechanism in the field of a heavy nucleus, which constitutes the primary SM background. In Section \ref{sec4}, we formulate the BSM cross sections for generic scalar, pseudoscalar, vector, and axial vector mediators, whereas Section \ref{sec5} presents the sensitivity projections and the optimization of the experimental setup to maximize signal significance. Finally, Section \ref{sec6} provides a summary of our work.

\section{MAGIX@MESA setup and apparatus}\label{sec2}

The MESA facility at Johannes Gutenberg University Mainz is designed to support a broad precision physics program at the intensity frontier \cite{Schlimme:2024eky}. The P2 experiment aims to measure the weak mixing angle, $\sin^2\theta_W$, at very low momentum transfer via parity-violating electron–proton scattering, providing a stringent test of the SM via electroweak radiative corrections \cite{Becker:2018ggl}, while the DarkMESA experiment plans to utilize the beam dump in the P2 experiment as a target and intends to search for signals of dark matter particles created in the dump \cite{MAGIX:2022fdt}. 

Complementing these is the MAGIX setup, which uses high-resolution magnetic spectrometers and internal gas targets to study electron scattering. Alongside precise determinations of proton electromagnetic form factors, it serves as a highly sensitive probe for weakly coupled light particles, such as dark photons and ALPs.

MAGIX will ultimately operate in the energy-recovering mode with a windowless gas-jet target to achieve the highest luminosities and minimize window-induced background \cite{A1:2021njh}. However, initial operations will utilize a conventional solid fixed target in extracted beam mode. For the BSM search program, a $^{181}\text{Ta}$ target is planned; its large atomic number $Z=73$ enhances the production cross section for processes scaling with $Z^2$, maximizing the expected signal yield. In the first phase, the beam energy is foreseen to be $55\,\text{MeV}$ and will later be increased to $105\,\text{MeV}$. 

The spectrometers of the MAGIX experiment setup are mounted on rotatable platforms and can be positioned independently around the common interaction point located at the center of the scattering chamber. The spectrometer STAR is installed on the right-hand side of the beam pipe, which direction is understood as $0^\circ$, while PORT is located symmetrically on the left-hand side.

STAR is designed to cover scattering angles from $-45^\circ$ to $165^\circ$, excluding the gap between $-15^\circ$ and $15^\circ$ in the beam direction, while PORT provides coverage from $-15^\circ$ to $-165^\circ$. The experiment geometry imposes the minimal angular separation between the spectrometers to be $47^\circ$. Their half-opening angles in both the polar and azimuthal directions are planned to be $2.75^\circ$ degrees. 

\section{Bethe-Heitler pair production in the field of heavy nuclei}\label{sec3}

BSM searches based on the Bethe–Heitler process in fixed-target experiments have previously been discussed in the context of dark photon searches at the Mainz Microtron \cite{A1:2011yso,Merkel:2014avp}. The maximum beam energy of $855\,\text{MeV}$ allowed sensitivity to dark photon masses in the range $50 \, \text{MeV} \lesssim m_{A}' \lesssim 300 \, \text{MeV}$, testing kinetic mixing parameters down to $\varepsilon \sim 10^{-3}$. The experiment resulted at the time in one of the most stringent constraints on the kinetic mixing of a light vector mediator.

\begin{figure*}[t]
\centering
\includegraphics[width=0.8\linewidth]{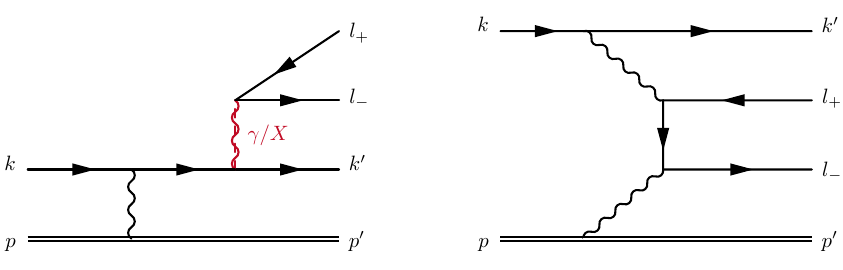}
\caption{Contribution to the Bethe-Heitler pair production, timelike (left) and spacelike (right) processes. The process on the left with a photon $\gamma$ replaced by the BSM mediator $X$ constitutes the main BSM background. Graphs which are obtained by crossing or anti-symmetrization are not shown.}
\label{fig:BackBH}
\end{figure*}

Since off-target radiation is suppressed due to the large nuclear mass, the relevant background contributions to such searches arise from the timelike and spacelike Bethe–Heitler amplitudes, illustrated in Fig.~\ref{fig:BackBH}. There, $k$ ($k'$) denotes the incoming (scattered) beam electron and $l_\pm$ the produced lepton pair. Four-momentum conservation is then expressed as

\begin{equation}
k + \Delta = k' + l_+ + l_-.
\end{equation} 
where the condition $\Delta_0 = 0$ is imposed, implying that the momentum transfer is purely spatial. We also introduce the following auxiliary variables

\begin{equation}
q = k-k', \quad \quad q' = l_++l_-,
\end{equation}
so that the two amplitudes can be factorized as

\begin{align}
\mathcal{M}_{\text{TL}} & = e^4  \frac{A^N_{\mu}I_{\text{TL}}^{\mu\alpha}j^{\text{pair}}_{\alpha}}{q'^2}, \\
\mathcal{M}_{\text{SL}} & = e^4 \frac{ A^N_{\mu}I_{\text{SL}}^{\mu\alpha}j^{\text{beam}}_{\alpha}}{q^2} ,
\end{align}
where the leptonic tensors are defined via

\begin{align}
I_{\text{TL}}^{\mu\alpha} = \bar{u}\left(k'\right)\gamma^{\mu} \frac{\slashed{k}-\slashed{q}' + m_e}{\left(k-q'\right)^2-m_e^2}\gamma^{\alpha} u\left(k\right) & + \text{crossed}, \\
I_{\text{SL}}^{\mu\alpha} =  \bar{u}\left(l_-\right) \gamma^{\mu} \frac{\slashed{q}-\slashed{l}_+ + m_\ell}{\left(q-l_+\right)^2-m_\ell^2}\gamma^{\alpha} v\left(l_+\right) & + \text{crossed},
\end{align}
and the associated leptonic currents are given by

\begin{align}
j^{\text{pair}}_{\alpha} & = \bar{u}\left(l_-\right) \gamma_{\alpha} v\left(l_+\right), \\
j^{\text{beam}}_{\alpha} & = \bar{u}\left(k'\right) \gamma_{\alpha} u\left(k\right). 
\end{align}

In the presence of two indistinguishable fermions in the final state, the total amplitude must be, in addition, antisymmetrized under their exchange, therefore taking the form

\begin{equation}\label{eq:antis}
\mathcal{M}^{\text{total}} = \mathcal{M}\left(k',l_-\right)-\mathcal{M}\left(l_-,k'\right),
\end{equation}
reflecting Fermi–Dirac statistics. This antisymmetrization is not required in processes such as muon-pair production, where the final-state particles are distinguishable. 

In the external-field approximation, the nuclear electromagnetic field is described by the static Coulomb potential in momentum space 

\begin{equation}\label{eq:NF}
A_{\mu}^N = \left(-\frac{Ze F\left(Q_t^2\right)}{\Delta^2}, 0,0,0\right),
\end{equation}
where $F\left(Q_t^2\right)$ is the nuclear form factor accounting for the finite spatial charge distribution, and we denoted $Q_t^2 \equiv -\Delta^2$. We also adopt the standard parameterization

\begin{equation}
\begin{split}
F\left(Q_t^2\right) &= 3\frac{e^{-S^2Q_t^2/2}}{\left(Q_t R_N\right)^3} \\
& \times\left[\sin{\left(Q_t R_N\right)} - \left(Q_t R_N\right)\cos{\left(Q_t R_N\right)}\right] ,
\end{split}
\end{equation}
which corresponds to a hard-sphere charge distribution of radius $R_N = 1.21 \tilde{A}^{1/3}  \, \text{fm}$, where $\tilde{A}$ is the nuclear mass number, supplemented by a Gaussian surface smearing characterized by $S = 0.9 \, \text{fm} $.

In the limit of an infinitely heavy nucleus, the differential cross section (using lab frame kinematics) reduces to the expression

\begin{equation}
d\sigma = \frac{1}{32}\frac{\left|\mathcal{M}\right|^2}{\left(2\pi\right)^8} \frac{\left|\mathbf{k}'\right|}{\left|\mathbf{k}\right|} dE_+dE_-d\Omega' d\Omega_-  d\phi_{\ell\ell} dm^2_{\ell\ell},
\end{equation}
where $E_\pm$ are the energies of the produced lepton pair, $\phi_{\ell\ell}$ is the azimuthal angle between them, $m^2_{\ell\ell}$ is their invariant mass squared, $\Omega'$ is the solid angle of the scattered electron, and $\Omega_-$ is the solid angle of the produced $l_-$ lepton. Further details on the integration over the final-state phase space, the kinematical setup, and the signal optimization are provided in Appendix \ref{appendix:a}.

\section{BSM contributions to the cross section}\label{sec4}

We examine the BSM contributions by considering generic scalar, pseudoscalar, vector, and axial-vector mediators. The corresponding interaction Lagrangian terms are given by
\begin{subequations}\label{eq:InrTerms}
\begin{align}
&\mathcal{L}_{S}= -\, g\, s\, \bar{\ell}\ell ,\\
&\mathcal{L}_{P}= -\, g\, a\, \bar{\ell}\gamma^5 \ell ,\\
&\mathcal{L}_{V} = -\, e\,\epsilon\, \bar{\ell}\,\slashed{V}\, \ell ,\\
&\mathcal{L}_{A} = -\, e\,\epsilon\, \bar{\ell}\,\gamma^{5}\slashed{A}'\, \ell,
\end{align}
\end{subequations}
where $s$ ($a$) is a scalar (pseudoscalar) field coupled to SM leptons with strength $g$, and $V_\mu$ ($A_\mu'$) denote the vector (axial-vector) mediator fields respectively, with $\epsilon$ originating from the kinetic mixing. 

Given the extremely small decay width, it is reasonable to apply the narrow width approximation
\begin{equation}
\frac{1}{\left(m^2_{\ell\ell}-m_X^2\right)^2+ \left(m_X \Gamma_X\right)^2} \to\frac{\pi\delta\left(m^2_{\ell\ell}-m_X^2\right)}{m_X \Gamma_X},
\end{equation}
where $m_X$ and $\Gamma_X$ denote the respective mediator mass and decay width. The latter is given by
\begin{equation}
\Gamma_X =\frac{\left| \mathcal{M}_{\text{decay}} \right|^2}{8\pi m_X^2} \sqrt{\frac{m_X^2}{4}-m_\ell^2},
\end{equation}
while the squared matrix elements for the scalar, pseudoscalar, vector, and axial-vector mediators are evaluated, respectively, as
\begin{equation}\label{eq:decayVA}
\left|\mathcal{M}_{\text{decay}} \right|^2 =  \begin{cases}
2g^2\left(m^2_s-4m_l^2\right), \\
2g^2m^2_a,   \\
4e^2\epsilon^2/3 \times\left(m_V^2+2m_\ell^2\right),\\
4e^2\epsilon^2/3 \times \left(m_A^2-4m_\ell^2\right).
\end{cases}
\end{equation}

In this scenario, only the timelike diagram contributes for the BSM process, as it alone yields the delta function responsible for a distinct resonance bump in the electron-positron pair spectrum. The effect of final-state antisymmetrization is washed out for the same reason, alongside the interference with the QED process.

To adapt the matrix element for these mediators, the following vertex replacements are sufficient
\begin{subequations}
\begin{align}
\text{scalar:} \quad & \gamma_{\alpha} \to 1, \\
\text{pseudoscalar:} \quad &  \gamma_{\alpha} \to \gamma^{5}, \\
\text{vector:} \quad & \text{vertex does not change},\\
\text{pseudovector:} \quad & \gamma_{\alpha} \to  \gamma_{\alpha}\gamma^{5}. 
\end{align}
\end{subequations}

Additionally, the coupling $e^2$ is replaced by $g^2$ for the scalar and pseudoscalar, and by $e^2\epsilon^2$ for the vector and axial vector. The photon propagator must also be replaced with the respective BSM particle propagator, incorporating the correct numerator, mass, and decay width.

In the kinematic region of interest, the only accessible decay modes are $e^+e^-$ and $\gamma\gamma$ pairs. While the Landau-Yang theorem forbids massive spin-1 particles from decaying into two photons, this channel remains open for scalars and pseudoscalars. However, the corresponding decay width scales as $m_X^3$, in contrast to the leptonic channel, which scales linearly with $m_X$. In the relatively low-mass range studied here, the diphoton mode is therefore strongly suppressed. Under the assumption of exclusively visible decays, neglecting this contribution is well justified.

\section{Sensitivity projections for BSM scenarios}\label{sec5}

For a gaussian distributed signal the sensitivity can be estimated as

\begin{equation}\label{eq:SB}
\frac{\sigma_{\text{BSM}}}{\sigma_{\text{background}}} = \frac{N}{\sqrt{L \times \sigma_{\text{background}}}},
\end{equation}
where $L$ denotes the integrated luminosity and $N$ represents the significance level (in standard deviations) required to distinguish a signal from background fluctuations. Conventionally, we set $N = 2$, corresponding to a 95\% confidence level\footnote{This analysis is purely statistical and no systematic errors are taken into account.}.

As seen from this equation, the reach for the BSM coupling scales as $L^{-1/4}$. Consequently, without a tailored optimization strategy, achieving a twofold improvement in coupling sensitivity requires a sixteen-fold increase in the collected dataset.

\begin{table}[t]
\centering
\setlength{\tabcolsep}{16pt}
\renewcommand{\arraystretch}{1.15}
\begin{tabular}{c|cc}
\hline\hline
Setup & PORT & STAR  \\
\hline
I  & $-45^\circ$ & $15^\circ$  \\
II & $-30^\circ$ & $30^\circ$  \\
\hline\hline
\end{tabular}
\caption{The benchmark setups which are used in the studies for the MAGIX@MESA experiment.}
\label{tab:SetupsMESA}
\end{table}

For the vector and axial-vector contributions, the timelike QED diagram depicted in Fig. \ref{fig:BackBH} constitutes an irreducible background, as it exhibits an identical angular distribution to the signal. It is therefore advantageous to maximize the overall event rate. Although this approach increases both the signal and background yields, the sensitivity reach scales with $\sqrt{\sigma_{\text{QED}}}$; thus, the overall statistical significance of the signal ultimately improves.

To achieve this goal, the STAR spectrometer is positioned at an angle of $15^\circ$ relative to the beam axis, while the PORT spectrometer is located at $-45^\circ$ (where the negative sign indicates an azimuthal angle of $\phi=180^\circ$ in spherical coordinates). By detecting the electron in PORT and the positron in STAR, this asymmetric configuration avoids the beam-induced pollution, while still benefiting from the forward enhancement of the cross section and covering the sufficiently broad range of mediator masses. This choice will be referred to as Setup I.

An alternative is to place the spectrometers symmetrically at $30^\circ$ and $-30^\circ$. This configuration, denoted as Setup II, yields constraints of comparable strength and is slightly more sensitive to higher-mass mediators. A summary of the corresponding parameters is provided in Table \ref{tab:SetupsMESA} for both setups. The energy selection threshold of $5 \, \text{MeV}$ is also assumed, with no additional momentum cuts imposed on the leptons accepted by the spectrometers.

\begin{figure}
    \centering
    \includegraphics[width=1.0\linewidth]{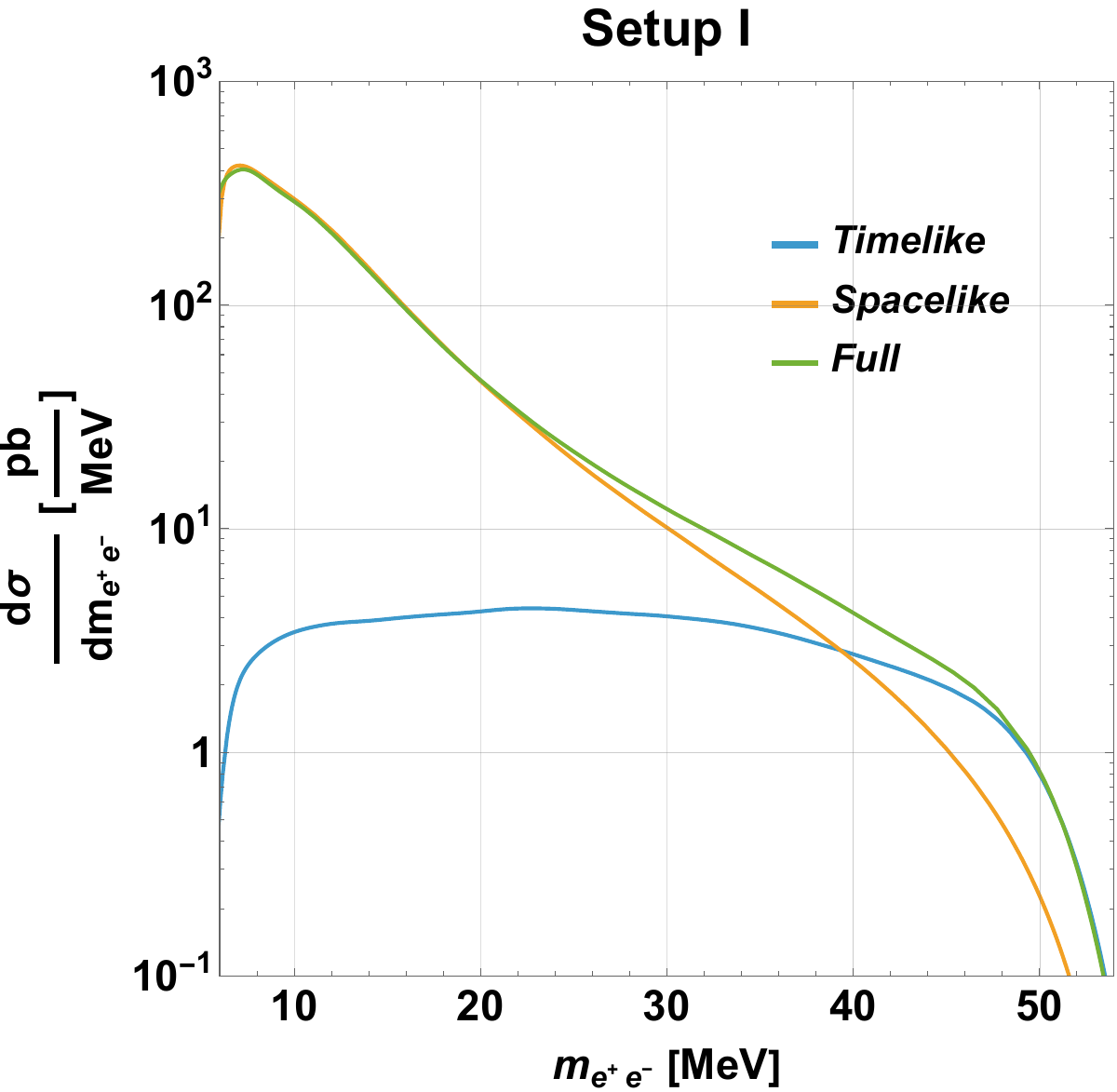}
    
    \vspace{12pt}

    \includegraphics[width=1.0\linewidth]{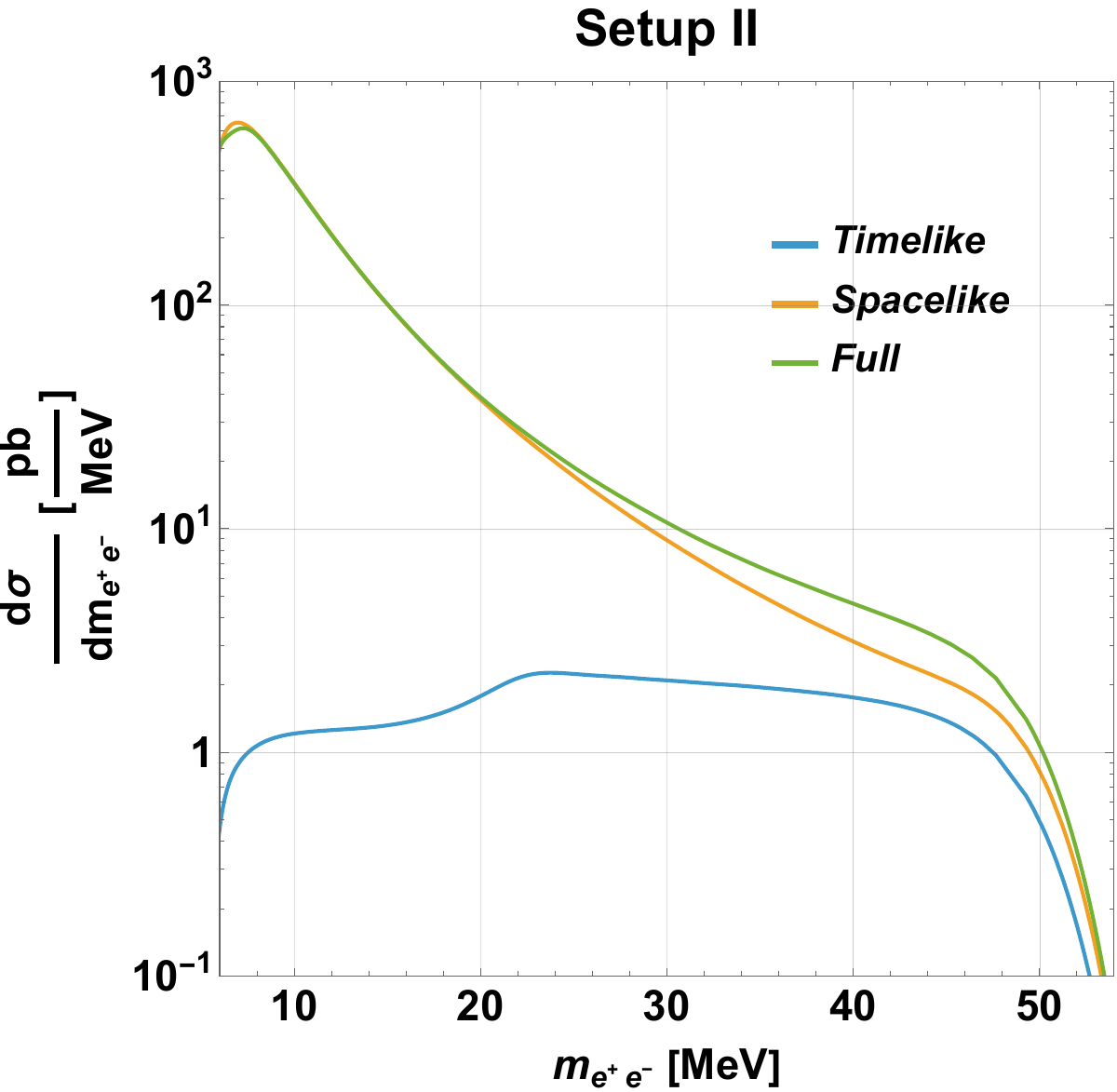}
    \caption{Differential distribution of the QED background with respect to the invariant mass of electron-positron pairs at the reference kinematics I and II (see Table \ref{tab:SetupsMESA}) and beam energy of $105 \, \text{MeV}$.}
    \label{fig:Background}
\end{figure}

Fig. \ref{fig:Background} illustrates the differential cross sections for the timelike and spacelike diagrams, along with their total contribution, at the beam energy of $105 \, \text{MeV}$. The spacelike background, which dominates at lower values of the electron-positron invariant mass, can be suppressed by selecting events where most of the beam energy is transferred to the produced pair. However, applying this kinematic cut substantially reduces the count rate and, given the fixed beam energy, restricts the accessible parameter space for the mediator mass, $m_X$. 

Before discussing the results, we address potential higher-order QED corrections to the Bethe-Heitler process, which may be sizable for a heavy tantalum target with $Z\alpha \approx 0.53$. While soft-photon radiative corrections have been studied for a proton target \cite{Heller:2021mcw}, the extension to a heavy nucleus remains to be done. Their effect, however, can be estimated. 
Since spacelike diagrams dominate across most of the $m_{e^+e^-}$ spectrum, we focus on their contribution. In a symmetric configuration, such as setup II, the $e^+e^-$ pair can be effectively treated as a neutral object which, in the first approximation, does not radiate or interact with the heavy nucleus, as the effect from electron and positron cancel each other. For an asymmetric configuration, such as setup I, the Feshbach formula \cite{McKinley:1948zz} indicates expected corrections at the level of $10\%$. 
However, the relation in Eq.~(\ref{eq:SB}) implies that the BSM coupling reach scales as $\sigma_{\text{background}}^{1/4}$. A $10\%$ variation in the background cross section therefore shifts the coupling sensitivity limits by less than $2.5\%$. Thus, the higher-order corrections may be safely neglected at first approximation for the purpose of BSM mediator searches\footnote[2]{A comparison between the bounds extracted from asymmetric setup I and symmetric setup II will, furthermore, allow to test this assumption.}.

\begin{figure*}
    \centering
    \includegraphics[width=0.43\linewidth]{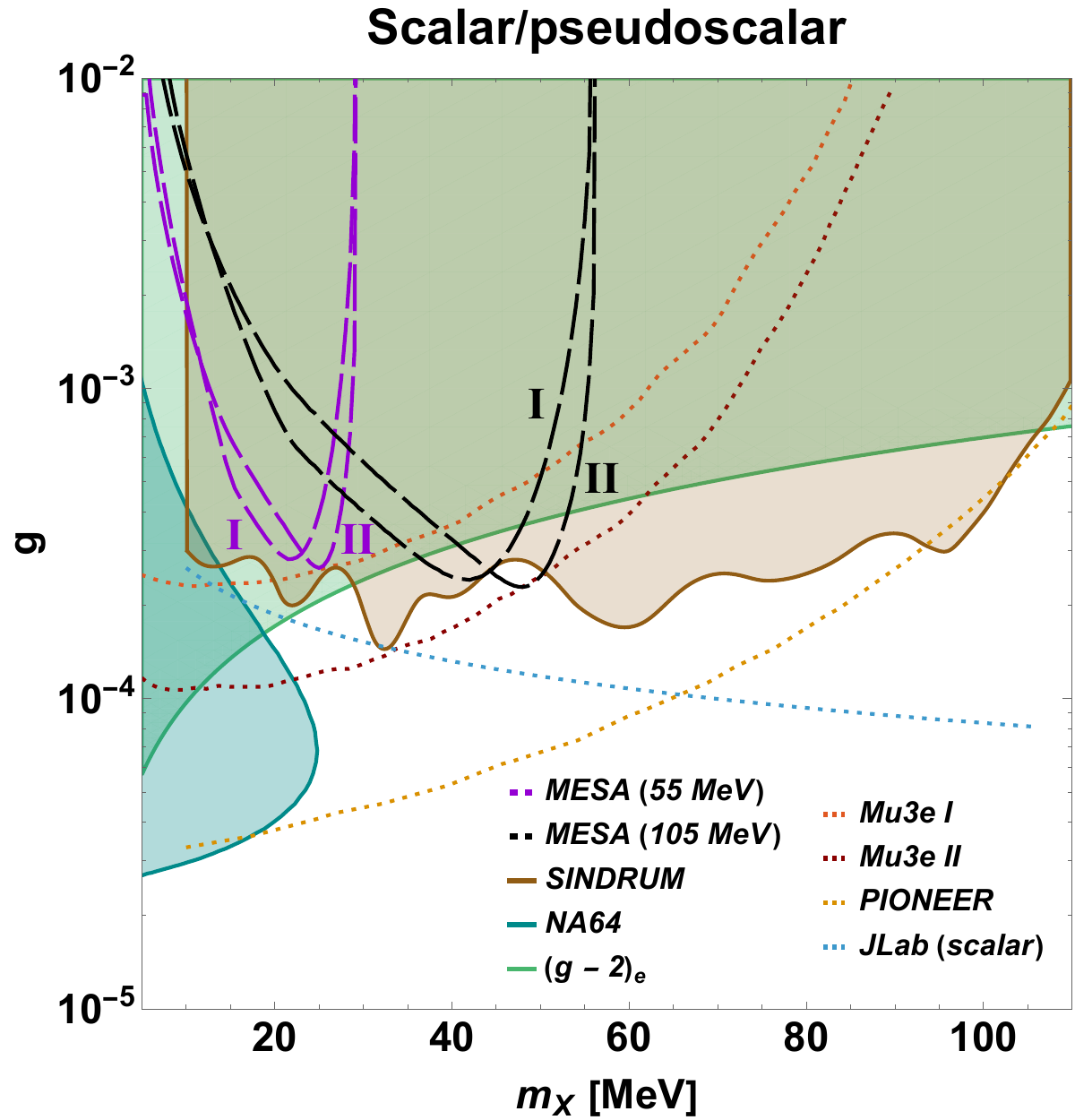} \quad \quad \quad
    \includegraphics[width=0.43\linewidth]{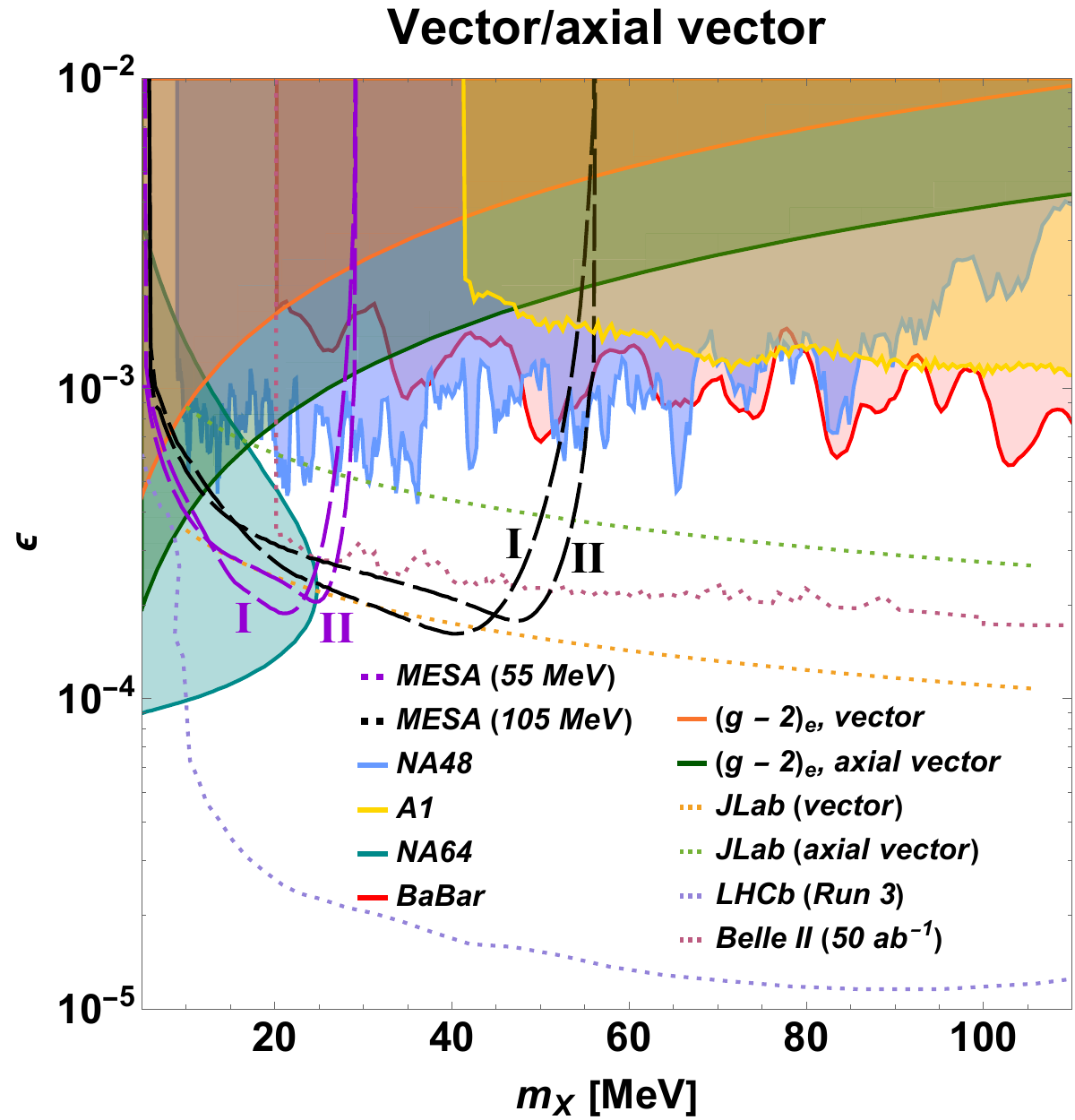}
    \caption{Constraints on the allowed BSM parameter space (value of coupling to electrons versus mediator mass) resulting from $\left(g-2\right)_e$ and from existing experiments searching for visible $e^+e^-$ decays (shaded exclusion regions), shown for the scalar/pseudoscalar (left) and the vector/axial vector scenarios (right). The projected sensitivity of the MAGIX@MESA experiment for two benchmark configurations, I and II (see Table \ref{tab:SetupsMESA}), is indicated by dashed lines, while other near-future experiments are shown as dotted lines and are described in the text. A single $\left(g-2\right)_e$ bound for scalar and pseudoscalar is shown, as the difference between them is negligible.}
    \label{fig:ExLimit}
\end{figure*}

Fig. \ref{fig:ExLimit} illustrates the projected sensitivity reach for scalar/pseudoscalar and vector/axial vector mediators at beam energies of 55 and 105 MeV derived from the bin-by-bin scan for a BSM contribution over the QED background. The bin width of $0.5 \, \text{MeV}$ is assumed for the first phase of experiment (at the beam energy of $55 \, \text{MeV}$) and improved to $0.1 \, \text{MeV}$ for the second stage (at the beam energy of $105 \, \text{MeV}$). Two weeks of continuous operation are assumed at the instantaneous luminosity  of $10^{35} \, \text{cm}^{-2} \text{s}^{-1}$.

In all cases the parity effects, i.e. the difference in the signal yield between scalar and pseudoscalar, as well as vector and axial-vector scenarios, were found to be negligible.

We compare these projections against existing bounds from $\left(g-2\right)_e$ measurements \cite{Fan:2022eto,Pustyntsev:2025nwm} and visible decay constraints on dark photons from A1@MAMI \cite{Merkel:2014avp}, BaBar \cite{BaBar:2014zli}, NA48/2 studies of $\pi^0$ decays \cite{NA482:2015wmo}, and NA64 analyses \cite{NA64:2019auh}. Additionally, we include projected sensitivities from the Belle II analysis \cite{Ferber:2015jzj}, the LHCb Run 3 projection based on dark photon radiation from charm mesons \cite{Ilten:2015hya}, and the JLab polarized positron beam program \cite{Pustyntsev:2025iht}.

For scalar and pseudoscalar mediators, we incorporate limits from the SINDRUM experiment \cite{SINDRUM:1989qan}, which searched for the $\pi^+ \to e^+\nu X$ process, as well as an adapted NA64 bound. The projected sensitivities include the PIONEER experiment, which aims to improve the branching ratio limit on the aforementioned pion decay to $\mathcal{O}\left(10^{-11}\right)$, and the Mu3e experiment. The Mu3e projections are based on the $\mu^+ \to e^+\bar{\nu}_\mu \nu_e X$ decay, assuming $2.5 \times 10^{15}$ and $5.5 \times 10^{16}$ produced $\mu^+$, respectively (labeled as I and II), adapted from \cite{DiLuzio:2025ojt}. In both the SINDRUM/PIONEER and Mu3e processes, the $X$ is emitted from the final-state positron and subsequently decays as $X \to e^+e^-$.

\section{Results and discussion}\label{sec6}

The sensitivity of MAGIX@MESA to visibly decaying BSM mediators was estimated assuming an effective data collection period of two weeks, utilizing a $0.5 \, \text{MeV}$ invariant mass bin width for the initial operational phase and a $0.1 \, \text{MeV}$ bin width for later stages (following the beam energy upgrade). These conservative baseline estimates leave room for further improvement under a dedicated search program. Despite this, the optimization of the experimental setup allows MAGIX@MESA to surpass the existing bounds. For vector and axial vector mediators, we demonstrate the capability to extend sensitivity reaches down to $\epsilon \sim 10^{-4}$. These bounds exceed the projected sensitivity of the Belle II experiment in this domain, assuming its end-goal dataset of $50 \, \text{ab}^{-1}$ integrated luminosity. Furthermore, they do not require additional assumptions regarding quark couplings and are complementary to the projected exclusion limits of the JLab polarized positron program.

In the case of scalar and pseudoscalar mediators, existing $\left(g-2\right)_e$ constraints appear more stringent than the projected MAGIX@MESA sensitivities. However, those indirect bounds are model-dependent and subject to potential Barr–Zee corrections \cite{Pustyntsev:2025nwm}. In contrast, MAGIX@MESA provides a clean, direct probe of the electron coupling where branching ratio effects induced by photon couplings remain negligible, as discussed above. We also emphasize that it represents one of the few experiments operating in this mass range capable of isolating electron–axion and electron–scalar couplings without relying on lepton universality or similar assumptions.

\section{Conclusions and outlook}

In this work, we have explored the sensitivity reach of the upcoming MAGIX experiment at the MESA facility for light, weakly-coupled BSM mediators in the sub-GeV mass range. By analyzing the production of generic scalar, pseudoscalar, vector, and axial-vector particles via the Bethe-Heitler process and their subsequent visible decay into $e^+e^-$ pairs, we demonstrated that the MAGIX setup provides a highly versatile probe of unexplored dark sector parameter space across various quantum numbers and coupling scenarios.

Specifically, the asymmetric double-spectrometer setup allows for an effective enhancement of the signal-to-background ratio. The studied benchmark spectrometer setup avoids beam-induced backgrounds while retaining high count rates in the forward direction, thereby maximizing the overall statistical significance of the BSM signal. We show that an effective data-taking period of two weeks on a heavy $^{181}\text{Ta}$ target is sufficient to yield highly competitive exclusion limits.

BSM couplings down to $\mathcal{O}\left(10^{-4}\right)$ can be probed for vector and axial-vector mediators, yielding exclusion limits that compete with or surpass the expected reach of other near-future experiments, such as Belle II and the JLab polarized positron program, in the corresponding mass range. Furthermore, for scalar and pseudoscalar mediators, MAGIX offers a robust, direct probe of the electron coupling, free of potential Barr--Zee corrections or other model-dependent uncertainties.

Finally, the results presented in this work correspond to the conventional heavy-target configuration of MAGIX@MESA, while its full physics potential will be realized only after the implementation of the energy-recovery mode with a gas-jet target. A detailed sensitivity study of this upgraded configuration would be highly valuable, but is left for future work.

\section{Acknowledgment}

The authors would like to thank Harald Merkel, Stefan Merkel and Sebastian Stengel for useful discussions. This work was supported by the Deutsche Forschungsgemeinschaft (DFG, German Research Foundation), in part through the Collaborative Research Center "CRC1660: Hadrons and Nuclei as discovery tools" (Project ID 514321794) and in part through the Cluster of Excellence “Precision Physics, Fundamental Interactions, and Structure of Matter” (PRISMA++ EXC 2118/2) within the German Excellence Strategy (Project ID 390831469).

\appendix
\section{Phase space integration}\label{appendix:a}

In this Appendix, we provide the technical details on the phase space integration, as well as the setup and optimization of the MAGIX@MESA experiment.

We begin by introducing an appropriate parameterization of the cross section. In the limit of an infinitely heavy nucleus, it reduces to the expression 

\begin{equation}
\begin{split}
& d\sigma = 2\pi \delta\left(E_i-E_f\right) \frac{\left|\mathcal{M}\right|^2}{2\left|\mathbf{k}\right|} \prod_{p\,=\,k', \,l_+, \, l_-} \frac{d^3 {\bf p}}{\left(2\pi\right)^32p_0}  \\
& = \frac{1}{16}\frac{\left|\mathcal{M}\right|^2}{\left(2\pi\right)^8} \frac{\left|\mathbf{k}'\right|\left|\mathbf{l}_+\right|\left|\mathbf{l}_-\right|}{\left|\mathbf{k}\right|} dE_+dE_-d\Omega' d\Omega_+ d\Omega_-,
\end{split}
\end{equation}

For the purposes of BSM studies it is convenient to parameterize the solid angles of final leptons as

\begin{equation}
d\Omega_+d\Omega_- = d\Omega_+d\Omega_{\ell\ell},
\end{equation}
where $\Omega_{\ell\ell}$ denotes the relative angle between $l_+$ and $l_-$. Using the standard relation between the relative angle and the pair invariant mass, we obtain

\begin{equation}
d\sigma = \frac{1}{32}\frac{\left|\mathcal{M}\right|^2}{\left(2\pi\right)^8} \frac{\left|\mathbf{k}'\right|}{\left|\mathbf{k}\right|} dE_+dE_-d\Omega' d\Omega_-  d\phi_{\ell\ell} dm^2_{\ell\ell}.
\end{equation}

Finally, two additional symmetry factors must be included. First, averaging over the helicity states of the initial electron introduces a factor of $1/2$. Second, because the two final-state leptons are identical and the amplitude is antisymmetrized with respect to their exchange, an additional combinatorial factor of $1/2$ is required. Together, these factors modify the overall normalization of the cross section to $1/128$.

We adopt a standard right-handed laboratory coordinate frame. The incident beam is aligned with the $z$-axis, and the STAR and PORT spectrometers are positioned to define  the horizontal $xz$-plane. 

Since the integration variables are transformed to the pair invariant mass $s_{\ell \ell}$, the positron's 3-momentum $\mathbf{l}_+$ is most naturally parameterized relative to the emitted electron. To achieve this, we construct a local orthonormal basis $\left(\mathbf{u}_x, \mathbf{u}_y, \mathbf{u}_z\right)$ directly aligned with $\mathbf{l}_-$.

First, we define the local longitudinal axis to point along the electron's direction
\begin{equation}
\mathbf{u}_z = \frac{\mathbf{l}_-}{|\mathbf{l}_-|}.
\end{equation}
Next, we define the local $y$-axis to be perpendicular to the plane spanned by the electron and the incident beam 
\begin{equation}
\mathbf{u}_y = \frac{\mathbf{k} \times \mathbf{l}_-}{\left|\mathbf{k} \times \mathbf{l}_-\right|}.
\end{equation}
The right-handed triad is then completed by defining the local $x$-axis
\begin{equation}
\mathbf{u}_x = \mathbf{u}_y \times \mathbf{u}_z.
\end{equation}
With this local basis established, the positron 3-vector is constructed by rotating it away from the electron axis by the pair opening angle $\theta_{\ell \ell}$ and the pair azimuthal angle $\phi_{\ell \ell}$
\begin{align}
 \mathbf{l}_+ &= \left|\mathbf{l}_+\right| \bigl( \sin\theta_{\ell \ell} \cos\phi_{\ell \ell} \, \mathbf{u}_x \nonumber \\
 &+ \sin\theta_{\ell \ell} \sin\phi_{\ell \ell} \, \mathbf{u}_y + \cos\theta_{\ell \ell} \, \mathbf{u}_z \bigr). 
\end{align}
So that the angle $\theta_{\ell \ell}$ can be expressed as
\begin{equation}
\cos \theta_{\ell \ell} = \frac{2m_{\ell \ell}^2+2E_+E_- - s_{\ell \ell}}{2 \left|\mathbf{l}_+\right|\left|\mathbf{l}_-\right|}.
\end{equation}

Finally, since most background counts originate from near-forward kinematics, where the initial electron scatters along the beam axis, escaping detection, it is considered lost, while the produced pair is assumed to be resolved within the spectrometers.

\bibliography{bibliography}

@article{Bertone:2004pz,
    author = "Bertone, Gianfranco and Hooper, Dan and Silk, Joseph",
    title1 = "{Particle dark matter: Evidence, candidates and constraints}",
    eprint1 = "hep-ph/0404175",
    archivePrefix1 = "arXiv",
    primaryClass = "hep-ph",
    doi = "10.1016/j.physrep.2004.08.031",
    journal = "Phys. Rept.",
    volume = "405",
    pages = "279--390",
    year = "2005"

}

@article{Planck:2018vyg,
    author = "Aghanim, N. and others",
    collaboration = "Planck",
    title1 = "{Planck 2018 results. VI. Cosmological parameters}",
    eprint1 = "1807.06209",
    archivePrefix = "arXiv",
    primaryClass = "astro-ph.CO",
    doi = "10.1051/0004-6361/201833910",
    journal = "Astron. Astrophys.",
    volume = "641",
    pages = "A6",
    year = "2020",
    note = "[Erratum: Astron.Astrophys. 652, C4 (2021)]"
}

@article{Krasznahorkay:2015iga,
    author = "Krasznahorkay, A. J. and others",
    title1 = "{Observation of Anomalous Internal Pair Creation in Be8 : A Possible Indication of a Light, Neutral Boson}",
    eprint1 = "1504.01527",
    archivePrefix = "arXiv",
    primaryClass = "nucl-ex",
    doi = "10.1103/PhysRevLett.116.042501",
    journal = "Phys. Rev. Lett.",
    volume = "116",
    number = "4",
    pages = "042501",
    year = "2016"
}

@article{Krasznahorkay:2021joi,
    author = "Krasznahorkay, A. J. and Csatl{\'o}s, M. and Csige, L. and Guly{\'a}s, J. and Krasznahorkay, A. and Nyak{\'o}, B. M. and Rajta, I. and Tim{\'a}r, J. and Vajda, I. and Sas, N. J.",
    title1 = "{New anomaly observed in He4 supports the existence of the hypothetical X17 particle}",
    eprint1 = "2104.10075",
    archivePrefix = "arXiv",
    primaryClass = "nucl-ex",
    doi = "10.1103/PhysRevC.104.044003",
    journal = "Phys. Rev. C",
    volume = "104",
    number = "4",
    pages = "044003",
    year = "2021"
}

@article{Holdom:1985ag,
    author = "Holdom, Bob",
    title1 = "{Two U(1)'s and Epsilon Charge Shifts}",
    reportNumber = "UTPT-85-30",
    doi = "10.1016/0370-2693(86)91377-8",
    journal = "Phys. Lett. B",
    volume = "166",
    pages = "196--198",
    year = "1986"
}

@article{Pospelov:2007mp,
    author = "Pospelov, Maxim and Ritz, Adam and Voloshin, Mikhail B.",
    title1 = "{Secluded WIMP Dark Matter}",
    eprint1 = "0711.4866",
    archivePrefix = "arXiv",
    primaryClass = "hep-ph",
    doi = "10.1016/j.physletb.2008.02.052",
    journal = "Phys. Lett. B",
    volume = "662",
    pages = "53--61",
    year = "2008"
}

@article{Ringwald:2012hr,
    author = "Ringwald, Andreas",
    title1 = "{Exploring the Role of Axions and Other WISPs in the Dark Universe}",
    eprint1 = "1210.5081",
    archivePrefix = "arXiv",
    primaryClass = "hep-ph",
    reportNumber = "DESY-12-170",
    doi = "10.1016/j.dark.2012.10.008",
    journal = "Phys. Dark Univ.",
    volume = "1",
    pages = "116--135",
    year = "2012"
}

@article{Baker:2013zta,
    author = "Baker, K. and others",
    title1 = "{The quest for axions and other new light particles}",
    eprint1 = "1306.2841",
    archivePrefix = "arXiv",
    primaryClass = "hep-ph",
    reportNumber = "DESY-13-089",
    doi = "10.1002/andp.201300727",
    journal = "Annalen Phys.",
    volume = "525",
    pages = "A93--A99",
    year = "2013"
}

@article{Bauer:2017ris,
    author = "Bauer, Martin and Neubert, Matthias and Thamm, Andrea",
    title1 = "{Collider Probes of Axion-Like Particles}",
    eprint1 = "1708.00443",
    archivePrefix = "arXiv",
    primaryClass = "hep-ph",
    reportNumber = "MITP-17-047",
    doi = "10.1007/JHEP12(2017)044",
    journal = "JHEP",
    volume = "12",
    pages = "044",
    year = "2017"
}

@article{Knapen:2017xzo,
    author = "Knapen, Simon and Lin, Tongyan and Zurek, Kathryn M.",
    title1 = "{Light Dark Matter: Models and Constraints}",
    eprint1 = "1709.07882",
    archivePrefix = "arXiv",
    primaryClass = "hep-ph",
    doi = "10.1103/PhysRevD.96.115021",
    journal = "Phys. Rev. D",
    volume = "96",
    number = "11",
    pages = "115021",
    year = "2017"
}

@misc{Fabbrichesi:2020wbt,
    author = "Fabbrichesi, Marco and Gabrielli, Emidio and Lanfranchi, Gaia",
    title = "{The Dark Photon}",
    eprint = "2005.01515",
    archivePrefix = "arXiv",
    primaryClass = "hep-ph",
    doi = "10.1007/978-3-030-62519-1",
    month = "5",
    year = "2020"
}

@article{Jaeckel:2010ni,
    author = "Jaeckel, Joerg and Ringwald, Andreas",
    title1 = "{The Low-Energy Frontier of Particle Physics}",
    eprint1 = "1002.0329",
    archivePrefix = "arXiv",
    primaryClass = "hep-ph",
    reportNumber = "CPT-10-18, DESY-10-016, IPPP-10-09",
    doi = "10.1146/annurev.nucl.012809.104433",
    journal = "Ann. Rev. Nucl. Part. Sci.",
    volume = "60",
    pages = "405--437",
    year = "2010"
}

@article{PADME:2025dla,
    author = "Bossi, F. and others",
    collaboration = "PADME",
    title1 = "{Search for a new 17 MeV resonance via e$^{+}$e$^{−}$ annihilation with the PADME experiment}",
    eprint1 = "2505.24797",
    archivePrefix = "arXiv",
    primaryClass = "hep-ex",
    doi = "10.1007/JHEP11(2025)007",
    journal = "JHEP",
    volume = "11",
    pages = "007 (2025)",
    year = "2025"
}

@article{Accardi:2020swt,
    author = "Accardi, A. and others",
    title1 = "{An experimental program with high duty-cycle polarized and unpolarized positron beams at Jefferson Lab}",
    eprint1 = "2007.15081",
    archivePrefix = "arXiv",
    primaryClass = "nucl-ex",
    doi = "10.1140/epja/s10050-021-00564-y",
    journal = "Eur. Phys. J. A",
    volume = "57",
    number = "8",
    pages = "261",
    year = "2021"
}

@article{Schlimme:2024eky,
    author = {Schlimme, S{\"o}ren and others},
    title1 = "{The MESA physics program}",
    eprint1 = "2402.01027",
    archivePrefix = "arXiv",
    primaryClass = "nucl-ex",
    doi = "10.1051/epjconf/202430306002",
    journal = "EPJ Web Conf.",
    volume = "303",
    pages = "06002",
    year = "2024"
}

@article{Becker:2018ggl,
    author = "Becker, Dominik and others",
    title1 = "{The P2 experiment}",
    eprint1 = "1802.04759",
    archivePrefix = "arXiv",
    primaryClass = "nucl-ex",
    doi = "10.1140/epja/i2018-12611-6",
    journal = "Eur. Phys. J. A",
    volume = "54",
    number = "11",
    pages = "208",
    year = "2018"
}

@article{MAGIX:2022fdt,
    author = "Christmann, Mirco and others",
    collaboration = "MAGIX",
    title1 = "{Light Dark Matter Searches with DarkMESA}",
    doi = "10.22323/1.398.0129",
    journal = "PoS",
    volume = "EPS-HEP2021",
    pages = "129",
    year = "2022"
}

@article{A1:2021njh,
    author = "Schlimme, B. S. and others",
    collaboration = "A1, MAGIX",
    title1 = "{Operation and characterization of a windowless gas jet target in high-intensity electron beams}",
    eprint1 = "2104.13503",
    archivePrefix = "arXiv",
    primaryClass = "physics.ins-det",
    doi = "10.1016/j.nima.2021.165668",
    journal = "Nucl. Instrum. Meth. A",
    volume = "1013",
    pages = "165668",
    year = "2021"
}

@article{A1:2011yso,
    author = "Merkel, H. and others",
    collaboration = "A1",
    title1 = "{Search for Light Gauge Bosons of the Dark Sector at the Mainz Microtron}",
    eprint1 = "1101.4091",
    archivePrefix = "arXiv",
    primaryClass = "nucl-ex",
    doi = "10.1103/PhysRevLett.106.251802",
    journal = "Phys. Rev. Lett.",
    volume = "106",
    pages = "251802",
    year = "2011"
}

@article{Merkel:2014avp,
    author = "Merkel, H. and others",
    title1 = "{Search at the Mainz Microtron for Light Massive Gauge Bosons Relevant for the Muon g-2 Anomaly}",
    eprint1 = "1404.5502",
    archivePrefix = "arXiv",
    primaryClass = "hep-ex",
    doi = "10.1103/PhysRevLett.112.221802",
    journal = "Phys. Rev. Lett.",
    volume = "112",
    number = "22",
    pages = "221802",
    year = "2014"
}

@article{McKinley:1948zz,
    author = "McKinley, William A. and Feshbach, Herman",
    title1 = "{The Coulomb Scattering of Relativistic Electrons by Nuclei}",
    doi = "10.1103/PhysRev.74.1759",
    journal = "Phys. Rev.",
    volume = "74",
    pages = "1759--1763",
    year = "1948"
}

@article{Heller:2021mcw,
    author = "Heller, Matthias and Keil, Niklas and Vanderhaeghen, Marc",
    title1 = "{Soft-photon radiative corrections to the $e^-p \to e^- pl^- l^+$ process}",
    eprint1 = "2108.02088",
    archivePrefix = "arXiv",
    primaryClass = "hep-ph",
    doi = "10.1103/PhysRevD.104.073007",
    journal = "Phys. Rev. D",
    volume = "104",
    number = "7",
    pages = "073007",
    year = "2021"
}

@article{Fan:2022eto,
    author = "Fan, X. and Myers, T. G. and Sukra, B. A. D. and Gabrielse, G.",
    title1 = "{Measurement of the Electron Magnetic Moment}",
    eprint1 = "2209.13084",
    archivePrefix = "arXiv",
    primaryClass = "physics.atom-ph",
    doi = "10.1103/PhysRevLett.130.071801",
    journal = "Phys. Rev. Lett.",
    volume = "130",
    number = "7",
    pages = "071801",
    year = "2023"
}

@article{Pustyntsev:2025nwm,
    author = "Pustyntsev, Aleksandr and Vanderhaeghen, Marc",
    title1 = "{Implications of recent (g-2){\ensuremath{\mu}} measurements for MeV-GeV dark sector searches}",
    eprint1 = "2506.17750",
    archivePrefix = "arXiv",
    primaryClass = "hep-ph",
    doi = "10.1103/b7wp-3vcj",
    journal = "Phys. Rev. D",
    volume = "112",
    number = "9",
    pages = "095001",
    year = "2025"
}

@article{BaBar:2014zli,
    author = "Lees, J. P. and others",
    collaboration = "BaBar",
    title1 = "{Search for a Dark Photon in $e^+e^-$ Collisions at BaBar}",
    eprint1 = "1406.2980",
    archivePrefix = "arXiv",
    primaryClass = "hep-ex",
    reportNumber = "BABAR-PUB-14-002, SLAC-PUB-15979",
    doi = "10.1103/PhysRevLett.113.201801",
    journal = "Phys. Rev. Lett.",
    volume = "113",
    number = "20",
    pages = "201801",
    year = "2014"
}

@article{NA482:2015wmo,
    author = "Batley, J. R. and others",
    collaboration = "NA48/2",
    title1 = "{Search for the dark photon in $\pi^0$ decays}",
    eprint1 = "1504.00607",
    archivePrefix = "arXiv",
    primaryClass = "hep-ex",
    reportNumber = "CERN-PH-EP-2015-093",
    doi = "10.1016/j.physletb.2015.04.068",
    journal = "Phys. Lett. B",
    volume = "746",
    pages = "178--185",
    year = "2015"
}

@article{NA64:2019auh,
    author = "Banerjee, D. and others",
    collaboration = "NA64",
    title1 = "{Improved limits on a hypothetical X(16.7) boson and a dark photon decaying into $e^+e^-$ pairs}",
    eprint1 = "1912.11389",
    archivePrefix = "arXiv",
    primaryClass = "hep-ex",
    reportNumber = "CERN-EP-2019-284",
    doi = "10.1103/PhysRevD.101.071101",
    journal = "Phys. Rev. D",
    volume = "101",
    number = "7",
    pages = "071101",
    year = "2020"
}

@article{Ferber:2015jzj,
    author = "Ferber, Torben",
    title1 = "{Towards First Physics at Belle II}",
    doi = "10.5506/APhysPolB.46.2285",
    journal = "Acta Phys. Polon. B",
    volume = "46",
    number = "11",
    pages = "2285",
    year = "2015"
}

@article{Ilten:2015hya,
    author = "Ilten, Philip and Thaler, Jesse and Williams, Mike and Xue, Wei",
    title1 = "{Dark photons from charm mesons at LHCb}",
    eprint1 = "1509.06765",
    archivePrefix = "arXiv",
    primaryClass = "hep-ph",
    reportNumber = "MIT-CTP-4702",
    doi = "10.1103/PhysRevD.92.115017",
    journal = "Phys. Rev. D",
    volume = "92",
    number = "11",
    pages = "115017",
    year = "2015"
}

@article{Pustyntsev:2025iht,
    author = "Pustyntsev, Aleksandr and Ramasamy, Muthubharathi S. and Vanderhaeghen, Marc",
    title1 = "{New physics searches via beam normal spin asymmetry in Bhabha scattering}",
    eprint1 = "2511.22568",
    archivePrefix = "arXiv",
    primaryClass = "hep-ph",
    doi = "10.1103/w2zp-7mdr",
    journal = "Phys. Rev. D",
    volume = "113",
    number = "7",
    pages = "075032",
    year = "2026"
}

@article{DiLuzio:2025ojt,
    author = "Di Luzio, Luca and Paradisi, Paride and Selimovic, Nudzeim",
    title1 = "{Hunting for a 17 MeV particle coupled to electrons}",
    eprint1 = "2504.14014",
    archivePrefix = "arXiv",
    primaryClass = "hep-ph",
    doi = "10.1016/j.nuclphysb.2025.117177",
    journal = "Nucl. Phys. B",
    volume = "1021",
    pages = "117177",
    year = "2025"
}

@article{SINDRUM:1989qan,
    author = "Egli, S. and others",
    collaboration = "SINDRUM",
    title1 = "{Measurement of the Decay $\pi^+ \to e^+ \nu_e e^+ e^-$ and Search for a Light Higgs Boson}",
    reportNumber = "PSI-PR-89-02",
    doi = "10.1016/0370-2693(89)90358-4",
    journal = "Phys. Lett. B",
    volume = "222",
    pages = "533--537",
    year = "1989"
}

\end{document}